\documentclass[12pt,preprint]{aastex}
\usepackage{epsfig}
\usepackage{rotating}
\usepackage[small]{caption}

\newcommand{\chan}{{\sl Chandra}}

\begin{document}

\title{Search for the Optical Counterpart of  the Vela Pulsar X-ray
Nebula\thanks{Based on
observations collected at the European Southern Observatory, La Silla,
Chile}
\thanks{Based on observations with the NASA/ESA Hubble Space Telescope,
obtained at the Space Telescope Science Institute, which is operated by
AURA,Inc.\ under contract No NAS 5-26555}
}

\author{R.\ P.\ Mignani}
\affil{European Southern Observatory, Karl-Schwarzschild-Str.\ 2,D85740, 
Garching, Germany}
\email{rmignani@eso.org}
\author{A.\ De Luca}
\affil{Istituto di Astrofisica Spaziale e Fisica Cosmica, Via Bassini 15,
I-20133 Milan, Italy} 
\affil{Universit\`a di Milano Bicocca, Dipartimento di Fisica, P.za della 
Scienza 3, I-20126 Milan, Italy} 
\author{O.\ Kargaltsev, G.\ G.\ Pavlov}
\affil{Pennsylvania State University, 525 Davey Lab, University Park, PA 16802, USA}
\author{S.\ Zaggia}
\affil{Osservatorio  Astronomico di  Trieste,  Via G.B.  Tiepolo 11,  I-34131
Trieste, Italy} 
\author{P.\ A.\ Caraveo}
\affil{Istituto di Astrofisica Spaziale e Fisica Cosmica, Via Bassini 15,
I-20133 Milan, Italy} 
\author{W.\ Becker}
\affil{Max Planck Institut f\"ur Extraterrestrische Physik, D-85748, 
Garching, Germany}

\begin{abstract}
Observations of  the Vela pulsar region with   the {\sl Chandra} X-ray
observatory have   revealed the  fine   structure of  its  synchrotron
pulsar-wind nebula (PWN), which showed  an overall similarity with the
Crab  PWN.  However, contrary  to the Crab,  no firm  detection of the
Vela PWN in optical has been reported yet.  To  search for the optical
counterpart of the X-ray  PWN, we  analyzed deep optical  observations
performed  with different telescopes.  We  compared the optical images
with  
those obtained with the   {\sl Chandra}  ACIS  to search for
extended emission patterns which could  be identified as  counterparts
of the X-ray nebula elements.  Although some features  are seen in the
optical  images, we find    no correlation with  the  X-ray structure.
Thus, we  conclude that the   diffuse optical emission is more  likely
associated with filaments  in the  host  Vela SNR.   The derived upper
limits  on the optical flux from  the PWN are compatibile, within the
uncertainties, with the values
expected on  the basis  of  the extrapolations of the  X-ray data.
\end{abstract}

\section{Introduction}

A compact  ($\sim 4'$) X-ray nebula around  the Vela  pulsar was first
 detected   in soft X-rays   by Harnden et   al.\ (1985) with the {\sl
 Einstein} High Resolution Imager (HRI).  These authors suggested that
 it is a  pulsar-wind nebula (PWN)  powered  by relativistic particles
 ejected by the  pulsar.  The soft X-ray  emission from the Vela X-ray
 nebula was  further  studied with {\sl  ROSAT}  (\"Ogelman, Finley \&
 Zimmerman 1993; Markwardt \& \"Ogelman 1998, and references therein).
 Based on its   ``kidney-bean'' shape, Markwardt  \&  \"Ogelman (1998)
 proposed  an interpretation in terms  of a bow-shock  produced by the
 supersonic motion of the pulsar through the ambient medium. 

The Vela  pulsar and its PWN have   been later observed with  both the
High Resolution Camera (HRC) and the Advanced CCD Imaging Spectrometer
(ACIS) on board  the {\sl Chandra} X-ray  observatory (Pavlov  et al.\
2000, 2001a,b, 2003; Helfand, Gotthelf, \& Halpern 2001; Kargaltsev et
al.\ 2002).  Thanks to the  excellent angular resolution of \chan, the
morphology  of the PWN was  resolved in a complex structure resembling
that of the Crab PWN. Such a structure cannot be explained by a simple
bow-shock model.  The  brighter part of the PWN,  with a size of about
$2'$, shows an approximately  axisymmetric structure, with two arcs, a
jet,  and a counter-jet, embedded  into  an extended diffuse emission.
The axis   of  symmetry,  which  can  be associated    with the pulsar
rotational axis  (Pavlov et al.\ 2000),  coincides, within the errors,
with the direction of the pulsar's proper motion (P.A.  = 301\degr ---
e.g., Caraveo et al.\ 2001a). 

The overall spectrum  of the PWN can be described by  a power law with
an  average   spectral  (energy)  index  $\alpha   \approx  0.5$  (see
Kargaltsev  et al.\  2002 for  details), which  can be  interpreted as
synchrotron emission of  relativistic electrons and/or positrons.  The
same electron distributions should  emit optical and radio synchrotron
radiation, provided that the  electron power-law spectrum extends down
to sufficiently  low energies, $\la  300 \nu_{14}^{1/2} B_{-4}^{-1/2}$
GeV (where $\nu = 10^{14} \nu_{14}$ Hz is the radiation frequency, and
$B=10^{-4}  B_{-4}$   G  is  the  magnetic   field),  and  synchrotron
self-absorption plays no role.

Indeed, highly  polarized ($\approx 60$\%   at 5.2 GHz)  extended radio
emission has  been recently detected around the  Vela pulsar (Lewis et
al.\ 2002; Dodson  et al.\ 2003),   covering a region  $\sim\!4$ times
larger than  the X-ray PWN, as observed  in the case  of the Crab.  
Most of the diffuse radio emission 
comes from two lobes
(see Figs.\ 2--4 in Dodson et al.\ 2003)  --- the southeast lobe of an
area  of 18 arcmin$^2$  ($F_\nu  = 760\pm 100$ mJy  at  5.2 GHz) and a
brighter northeast lobe of an area of 5.3  arcmin$^2$ ($F_\nu = 290\pm
50$ mJy at 5.2 GHz).  Search for radio  emission from the X-ray-bright
compact nebula was hampered by the brightness of the pulsar.  Although
some emission was detected (e.g, about 30 mJy at $\nu = 2.4$ GHz, in a
0.73 arcmin$^2$  area around the  pulsar),  it can well be  
an unsubtracted pulsar contribution
and should be considered as an upper limit on the compact
PWN emission.

One   could expect to  see   optical nebula   around the Vela  pulsar.
Indeed,  optical  PWNe have   been  observed in   several other  young
pulsars.  The most famous example  is the Crab  PWN (see, e.g., Hester
et  al.\ 2002,  and  references  therein)   which  exhibits a  complex
structure featuring bright  knots  and sharp arc-like  features dubbed
``wisps''  (Scargle 1969) located at  the inner  boundary of the X-ray
torus and probably associated with  a termination shock in the  pulsar
wind (e.g., Hester et  al.\  1995; Kennel  \& Coroniti 1984).  Also,
the  comparison between the {\sl
Hubble   Space Telescope} ({\sl   HST})   and \chan\ observations
of the LMC pulsar B0540--69  has
clearly shown the presence of an optical PWN which follows exactly the
morphology of the X-ray PWN (Caraveo et al.\ 2001b, 2003). 

Contrary to the Crab and  PSR B0540--69, searches for  the Vela PWN in
optical have been inconclusive so far.  The only marginal detection of
an optical PWN was reported by \"Ogelman et  al.\ (1989), who observed
the Vela pulsar field  in  the $V$ and $B$  bands  with the ESO  2.2 m
telescope.  These authors found  evidence for optical diffuse emission
(typical size $\sim 2'$), with  an average surface brightness of about
26 mag  arcsec$^{-2}$. However, the presence  of bright filaments from
the host SNR, as well as of several bright stars in the field, made it
difficult to assess the reality of the putative optical nebula. 
 
The   increase in sensitivity  and  angular resolution provided by the
{\sl HST} prompted us to carry out  a new optical investigation of the
Vela pulsar region.  In the following, we discuss spatial correlations
between {\sl Chandra}  ACIS images (Pavlov et   al.\ 2001b; Pavlov  et
al.\ 2003) and the optical ones collected 
by both the {\sl HST} WFPC2
(Mignani and Caraveo 2001; Caraveo et al.\ 2001a)  and 
ESO {\sl
NTT} (Nasuti et al.\ 1997) and {\sl VLT}. 

\section{X-ray and optical observations}

\subsection{Chandra observations}

The multiple  observations of  the Vela PWN  with the \chan\  ACIS are
described by Pavlov  et al.\ (2003).  In all  the observations the PWN
was imaged  on the  same back-illuminated chip  S3, more  sensitive to
soft X-rays.  Although the  individual images show some variability of
the  PWN elements,  the  overall PWN  morphology  remains stable.   To
obtain a very  deep X-ray image to be compared  with the optical ones,
we stacked  seven consecutive  individual images, collected  from 2001
November 25 through 2002 April 3  for a total exposure time of 141 ks,
and adaptively smoothed the combined image.  The accuracy of the image
co-alignment was about  0\farcs7.  The smoothed X-ray image  of the PWN,
in the energy band 1--8 keV, is shown in 
Figure 1. The logarithmic brightness  scale was chosen to increase the
dynamic range.   The jet and the counter-jet,  northwest and southeast
of the  pulsar, respectively, and two arc-like  structures (inner and outer arcs)
are the dominant features of the bright inner nebula (upper
panel).  The inner  nebula is 
surrounded by  a bean-shaped region
of diffuse emission (outer nebula), $\sim 2'\times 2' $ in size (lower
panel).  An elongated region of fainter extended emission southwest of
the 
bright inner PWN is  seen up to the edge of the ACIS image
($\sim  4'$  from the  pulsar).   A  $100''$-long  outer jet  is  seen
northwest of  the inner  PWN.  The spectrum  of the bright  inner PWN,
within a  $45''$ radius around the  pulsar, can be fitted  by a single
power-law  model  with  a  spectral index  $\alpha=0.47\pm  0.02$  and
unabsorbed  flux  of  $9.1\pm   0.7  \times  10^{-11}$  ergs  s$^{-1}$
cm$^{-2}$, in  the 0.1--10 keV  range (Kargaltsev et al.\  2002).  The
unabsorbed surface brightnesses for the outer and inner arcs are about
$6 \times  10^{-14}$ and $3  \times 10^{-14}$ ergs  s$^{-1}$ cm$^{-2}$
arcsec$^{-2}$, while  the spectral indices $\alpha$ are  about 0.4 and
0.3, respectively.

\subsection{Optical wide-band imaging}

Optical observations of   the  Vela pulsar  field were   collected  at
different    epochs    using  different telescopes,   instrumentation,
observational set-ups, and     filters.  In addition   to  the  recent
high-resolution imaging data  obtained with the  {\sl HST},  our data
set includes  archived ground-based images  obtained from ESO with the
{\sl NTT} and the {\sl VLT}.  Table 1 gives a  detailed summary of the
available observations. 

\subsubsection{HST}

{\sl HST} observations of the Vela pulsar field were collected between
1997 June and  2000 March with  the WFPC2.   Several images have  been
taken  through    the   filters   $675W$   ($\lambda=6717$\AA;   $\Delta
\lambda=1536$\AA),   $814W$ ($\lambda=7995$\AA;   $\Delta \lambda      =
1292$\AA) and $555W$  ($\lambda=  5500$\AA; $\Delta \lambda  = 1200$\AA)
with exposure times ranging between 2000 s and 2600 s. These data were
originally  taken  as part of  two  independent programs aimed  at the
study  of the multicolor flux  distribution of  the pulsar (Mignani \&
Caraveo  2001) and   at the measure   of  its parallactic displacement
(Caraveo et al.\ 2001a).   To achieve the maximum spatial  resolution,
in all the exposures  the pulsar  was located  near the center  of the
Planetary Camera  chip, with a pixel size  of $0\farcs045$ and a field
of view of $35''  \times 35''$.  The standard  {\sl HST} pipeline  was
applied  for  the image   reduction    and for  the 
photometric calibrations.  Co-aligned  exposures were finally combined
to filter out cosmic ray hits (see Mignani \& Caraveo 2001 and Caraveo
et al.\ 2001a for a detailed description of  the observations and data
reduction). 

For each  observation, the IRAF/STSDAS task  {\tt wmosaic} was used to
obtain  a  mosaic of  the   four WFPC2  images    and correct for  the
instrumental geometric distortion  of the cameras (Casertano \&  Wiggs
2001).   To increase the   signal-to-noise ratio, we combined  all the
 available $555W$  filter   images.   Since the   observations  at
different epochs were performed with different telescope pointings and
roll  angles, to superpose  the frames we  had to follow the procedure
successfully applied in previous astrometric works, 
using a common
reference frame defined by a grid of reference stars' coordinates (see
Caraveo    et al.\   2001a      and references  therein   for   further
details). Images  were  registered   within $\approx0.1$   WFC  pixels
($\approx$  10 mas),  an accuracy adequate  to  achieve  the goals of  the
present investigation, co-aligned and stacked.  

\subsubsection{NTT}

The  Vela pulsar  field was  observed  with the  {\sl NTT} at  the La  Silla
observatory  in January  1995 (Nasuti  et al.\  1997). To  achieve the
highest sensitivity  both at longer and shorter  wavelengths, the EMMI
camera was used in both its  Red (EMMI-R) and Blue (EMMI-B) arm modes.
The camera is a TEK $2048\times  2048$ pixels CCD with a pixel size of
0\farcs27 (field  of view $9\farcm2  \times 8\farcm6$ )  and 0\farcs37
(field of  view $6\farcm2 \times  6\farcm2$) in the EMMI-R  and EMMI-B
configuration  modes,  respectively.   Images  were collected  in  the
Johnson's   $U$  ($\lambda=3542$\AA;  $\Delta   \lambda=542$\AA),  $B$
($\lambda=4223$\AA; $\Delta  \lambda=941$\AA), $V$ ($\lambda=5426$\AA;
$\Delta   \lambda=1044$\AA),  and   $R$   ($\lambda=6410$\AA;  $\Delta
\lambda=1540$\AA) bands with exposure times between 15 and 40 minutes.
Data  reduction  and  photometric  calibration  were  carried  out  as
described in Nasuti et al.\ (1997).

\subsubsection{VLT}

Observations of the field were performed in  April 1999 with the first
Unit Telescope   (Antu) of the ESO  {\sl  VLT} located  at the Paranal
Observatory.  Images were    obtained  using the   FOcal  Reducer  and
Spectrograph \#1 (FORS1)  instrument, a four-port  $2048 \times 2048 $
CCD detector   which  can  be   used  both as   a  high/low resolution
spectrograph and  an imaging camera.   The instrument  was operated in
imaging mode  with a $1\times1$  binning  and at  its standard angular
resolution of 0.2 arcsec/pixel, with  a corresponding field of view of
$6\farcm8 \times 6\farcm8$.  Two  300 s  exposures  were taken in  the
Bessel filters  $R$  ($\lambda=6570$\AA; $\Delta\lambda=1500$\AA)  and
$I$   ($\lambda=7680$\AA; $\Delta\lambda=1380$\AA)  under good  seeing
conditions $(\le 1''$).  The images have been  retrieved from the  ESO
public archive    and  reduced   following standard   procedures   for
debiassing and flatfielding.   Since no standard stars were  observed,
absolute flux calibration was performed using as a  reference a set of
secondary stars  detected in  the   same passbands in the  {\sl   NTT}
images.

\subsection{Narrow-band imaging}

Three 20-min exposures of the
Vela pulsar   field 
were taken in  H$_\alpha$ ($\lambda   =
6588.27$\AA, $\Delta \lambda=74.31$\AA) on 1999 April  4 with the Wide
Field Imager  (WFI) at the ESO/MPG 2.2~m  telescope. The WFI is a wide
field mosaic camera, composed   of eight $2048\times4096$  pixel CCDs,
with a scale  of  $0.238$ arcsec/pixel,   
providing a field of view  of
$33\farcm7\times 32\farcm7$.  
To compensate for the
loss of  signal  due to the  interchip  gaps (23\farcs8 and  14\farcs3
along right ascension  and declination, respectively), the  second and
third exposures were taken with a relative offset of $30\arcsec$ in RA and Dec.  The images  were taken in fairly good
seeing conditions with  FWHM$\simeq0\farcs9$.  Data reduction with the
usual CCD  processing  steps of  bias   subtraction, flat-fielding and
trimming   of  images,    was   performed with  IRAF\footnote{IRAF  is
distributed by  NOAO,  which  is operated   by  the AURA, Inc.,  under
cooperative agreement with the  NSF.}  using the Mosaic CCD  reduction
package (MSCRED).  Individual  exposures have been finally coadded and
cleaned of  cosmic-ray  hits using   the   DRIZZLE software (Hook   \&
Fruchter 1997) and the final image  has been corrected for the effects
of fringing that  affects the  CCDs in the  red part  of the  spectrum
($\gtrsim6500$ \AA).

\subsection{Image superposition: X-ray versus optical}

The direct comparison of images taken in different energy bands (e.g.,
optical   and  X-rays)   can  be   achieved  through   accurate  image
superposition.  Following the approach used by Caraveo et al.\ (2001b),
we superimposed  the X-ray {\sl  Chandra} image onto the  optical ones
with  respect  to the  absolute  ($\alpha$,$\delta$) reference  frame,
relying  on the  astrometric  solution  of each  image.   We used  the
combined  {\sl  Chandra}  ACIS  image, where  individual  images  were
co-aligned with an accuracy of  $0\farcs7$ (Pavlov et al.\ 2003).  The
error of the absolute  {\sl Chandra} aspect solution\footnote{see {\tt
http://cxc.harvard.edu/cal/ASPECT/celmon/}}    is    typically   about
$0\farcs6$.  Therefore,  we adopt $1''$  as a reasonable  estimate for
the uncertainty of the absolute X-ray astrometry.

For the WFPC2   images, the default   astrometric solution across  the
focal plane  is derived from  the  coordinates of the  two guide stars
used to point the telescope, which are taken as a reference to compute
the  astrometric reference point  and the telescope   roll angle.  The
accuracy of the  {\sl HST}  astrometry  is 
limited
 by the  intrinsic
error on the  absolute coordinates of  the GSC1.1 (Guide Star Catalog)
stars (Lasker et al.\ 1990) which are used for the telescope pointing.
According to the   current  estimates,  the mean uncertainty   of  the
absolute positions    quoted  in the   GSC1.1  is  about  0\farcs8 per
coordinate  (Biretta  et al.\ 2002).    { For this reason,  we have
recomputed a new astrometric solution for the WFPC2 images by using as
a reference the positions of several  stars from the USNO-A2.0 catalog
(Monet 1998)    identified in   the    field of  view.   The    ASTROM
software\footnote{http://star-www.rl.ac.uk/Software/software.htm}  was
used to compute  the pixel-to-sky coordinates transformation. The  rms
of the astrometric fit was $\sim 0\farcs08$, per coordinate. 

The astrometric calibrations  of  the the ground-based  optical images
was computed  as  described above.   The astrometric fits  yielded rms
values of $\sim 0\farcs13$ and $\sim 0\farcs22$ per coordinate for the
{\sl VLT} and {\sl  NTT} images, respectively.  For the WFI image, the
astrometry was  computed separately for  each of the eight  CCD chips,
yielding an average rms of $\sim 0\farcs3$. 

In  all cases, the final  uncertainty on our astrometric solutions was
evaluated by accounting both for the rms of  the astrometric fits and for
the propagation of the error on the intrinsic absolute accuracy of the
USNO-A2.0     coordinates,   which       is  of     the    order    of
$0\farcs2$--$0\farcs3$. 

The  absolute frame registration between  the optical and X-ray images
turned out  to be  accurate within  {$\approx 1''$}, i.e.,  compatible
with the overall   uncertainties of the  absolute astrometry  of  each
frame.  Unfortunately, the 
strong pile-up  of the Vela pulsar in the
ACIS images 
hampered the use of the pulsar position
for better co-alignment
between the X-ray and optical images.

\section{Results}

\subsection{Search for a compact optical nebula}

As  a   first  step,  the  central   part  of  the   X-ray  PWN  field
(corresponding to the inner nebula and the outer nebula) was inspected
to search for  optical counterparts of the complex structures
seen  
with {\sl  Chandra}.  Our  starting point  was the  combined WFPC2
$555W$ image,  which is by far  the deepest optical image  of the Vela
pulsar  field.   The  final image  is  shown  in  Figure 2,  where  we
superimposed  the X-ray contour  map obtained  from the  combined {\sl
Chandra}  ACIS exposure  of the  region.    Although  a number  of
complicated patterns of diffuse emission are present in the $\approx 3'
\times 3'$ 
field of view, no optical  counterparts of the X-ray
features seen in  Figure 1 can be identified,  nor any other structure
symmetric with respect to the axis of symmetry  of the 
X-ray PWN. We thus conclude that the
PWN is undetected in the optical.

To compute 
the upper  limit on the optical surface brightness of
the PWN, an accurate mapping of the background 
is required.  This 
is complicated by the presence of diffuse, non-uniform, emission patterns
which show  sharp surface brightness  variations on angular  scales as
small  as $\approx  5$  arcsec.   For this  reason,  we evaluated  the
background level  in $\approx200$ cells of 1  arcsec$^2$ each, selected
in a number of star-free  regions across the whole image.  Statistical
errors on  the number of counts per  cell were of the  order of 0.3\%.
The background level, 
$\approx 22.3$ ST 
magnitudes\footnote{http://www.stsci.edu/instruments/wfpc2/wfpc2\_doc.html}
~arcsec$^{-2}$ on average,
was found to vary typically 
by 4\%--5\% across the
whole field, with a maximum variation of $\approx$ 7\%. 

For an extended  source, the  upper limits  on the surface  brightness
scale with the detection area ${\cal A}$ as ${\cal  A}^{-1/2}$.
An optimally chosen area should be 
large enough  to reduce the statistical  errors  due to the background
fluctuations but it should be smaller
than the scale  of
sharp   background  variations.  According    to  our mapping  of  the
background, a detection area of 10 arcsec$^{2}$ represents
a resonable  optimization.  

For the 
area chosen,
the measured background variations 
affect the  
upper limit on the flux of an 
  extended source
(at a 3 $\sigma$ level) by  no more
than $\approx$3\%, corresponding to surface brightness variations 
below 0.1 magnitudes arcsec$^{-2}$.

While variations in  the sky background across the  image play a minor
role,  we found that the  derived surface  brightness upper limits are
different in different regions of  the image.  This  is mainly due  to
the non-uniform coverage of the field performed by the WFPC2. 
First,
because  of the intrinsic differences  in the pixel  size and in the
physical characteristics of the CCDs, the PC and the WFC chips
contribute differently  to the instrumental  background per unit area.
In addition, since the five  WFPC2 $555W$ observations listed in Table
1 were executed at different epochs  and with different telescope roll
angles, the exposure map  varies across the  field.  For instance, the
central region  of the field, which is  covered by the PC ($35''\times
35''$), reaches  an integration  time of  11\,800 s, while  the  outer
regions,  covered   by the three  WFC  chips,  have  integration times
varying between 4\,600 and 11\,800 s. 
 
Both  effects   clearly  
affect the  evaluation   of  the  surface
brightness upper limits  in different regions of the  PWN, as they are
covered  differently by  the four  WFPC2 chips.   As it  is  seen from
Figure  2,  some regions  of  the inner  nebula  (inner  arc, jet,  and
counter-jet)  fall entirely  within the  PC,  while the  outer arc  is
coincident with the inter-chip gaps and is covered partially by the PC
and partially by  the WFC chips.  On the other  hand, the outer nebula
is entirely covered by the WFC chips.

Taking all these effects into account, we  have computed the 3$\sigma$
upper  limits  on the  optical surface  brightness  of both  the inner
nebula (inner arc, jet, counter-jet, outer  arc) and the outer nebula.
We note that although our upper  limits have been  computed 
for a detection area of 10 arcsec$^{2}$, they
can be easily rescaled to any other 
area.

Using the {\sl HST}  pipeline photometric calibration, we 
computed
an    upper     limit    of    
28.1 and 28.0--28.5
ST
magnitudes  arcsec$^{-2}$ for  the inner  nebula  and for  the outer  nebula,
respectively.  As  we mentioned before, because  of the non-uniformity
of the exposure map the upper limit for the outer nebula turned out to
be slightly 
position-dependent.
To correct these limits for the interstellar reddening, 
we use the extinction $A_V=0.2$, consistent with the hydrogen column
density estimated from the X-ray observations of the Vela pulsar
(Pavlov et al.\ 2001a). The corrected limits are 
27.9 and 27.8--28.3 ST magnitudes arcsec$^{-2}$.

The  same analysis  was then  repeated for  the other  available WFPC2
images ($675W$ and  $814W$) as well as for the  {\sl NTT} ($UBVR$) and
{\sl VLT}  ($VI$) ones  (see Table  1) but no  evidence for  a compact
optical nebula  was found in  either of these datasets.   The computed
upper    limits   are    summarized   in    Table   2.

\subsection{Search for an extended nebula}
\label{extended}

Since the emission  from the  PWN could, in  principle,  be visible at
optical wavelenghts  on  larger angular  scales  with respect  to  the
X-rays, as it has been observed in radio (see \S 1), we took advantage
of  the  larger  field  of  view provided  by   the {\sl NTT}  images
to  search for  extended
features up  to distances of  $\approx 3$  arcmin.  In particular,  we
searched for diffuse optical emission at the position of the southwest
extension of the X-ray nebula (see Fig.\ 1, lower panel).  To  go as deep as
possible, we  have combined all  the  available {\sl NTT}  $UBVR$ band
images (Fig.\ 3,  upper panel).    The combined image shows  many  different
enhancements   in the   background, with  a   rather complex   spatial
distribution.  However, none of them can be firmly correlated with the
known X-ray   features.  In  addition,   we note that  almost  all the
diffuse emission  patterns seen in  the {\sl NTT}  $UBVR$ image can be
also identified in the  ESO/2.2m H$_\alpha$  (Fig.\ 3, lower  panel).
This suggests that  they are most   likely associated with  the bright
filaments of  the  Vela supernova  remnant.    We   note that  the
computed {\sl NTT} upper limits 
on the optical emission of the compact
X-ray nebula 
can  be applied  also at larger  distances
from   the pulsar.  The maximum   variation  in surface brightness for
source-free  regions, due to  the  complicated distribution of diffuse
emission, is found to be of order 7\% even in the outer regions of the
{\sl NTT} field,
which corresponds to $<0.1$ magnitudes arcsec$^{-2}$ variations of the upper
limit across the field (see \S 3.1).   
We note that for the {\sl NTT} images the measured upper limits in the
$U$ and  $B$ bands do  not apply to  the  region $\approx$  2.5 arcmin
southwest  of the   pulsar position, 
close to the
southwest edge of  the  extended  X-ray nebula.  
The background in this region is unrecoverably polluted by the presence of
a bright ($B\approx$9) B star that is not visible  in the $V$ and $R$
band images because of the slightly narrower field of view.

}

\section{Discussion}

It is interesting to compare  the measured upper limits on the optical
brightness of the Vela PWN with the X-ray and radio observations.  The
observed X-ray  spectra of the Vela  PWN are described by  a power law
with  spectral indices  $\alpha \approx  0.3$--0.4  and $\alpha\approx
0.4$--0.5, for the  
arcs and the extended
diffuse emission southwest of the pulsar, respectively (Kargaltsev et
al.\ 2002).  The X-ray  and radio surface brightness spectra, together
with  the  optical  upper  limits,  are  shown in  Figure  4  for  the
outer arc (upper panel) 
and  the diffuse emission  southwest of
the pulsar (lower panel).
Since the  inner PWN structures
were not resolved in the radio (see \S1), only radio 
upper limits are
plotted in  the upper panel.   
From the plot  we see that both the optical
and radio upper limits for  the outer arc are within  the
uncertainty of the extrapolation of the  X-ray spectrum towards  lower frequencies.


In  the case  of  the southwest  diffuse  emission, we  note that  the
X-ray-to-radio extrapolation  is well  {\em below} the  measured radio
brightness values.   This apparent  inconsistency can be  explained by
the  fact that  the  radio  and X-ray  brightnesses  were measured  in
different  areas (the X-ray  image is  substantially smaller  than the
radio image).  Moreover, even  within the smaller X-ray field-of-view,
it is seen that the X-ray brightness of the southwest diffuse emission
fades towards the region of maximum radio brightness (which is outside
the X-ray field of view).  Such behavior can be explained by radiative
(synchrotron)  and  adiabatic  cooling   of  the  expanding  cloud  of
relativistic electrons, which  can result not only in  the increase of
the spectral index, but also in the shift of the lower-energy boundary
of the  power-law spectrum towards lower energies.   Therefore, we can
expect  maximum  optical brightness  to  be  observed  in between  the
regions  of the X-ray  and radio  maximum brightnesses.   Although the
X-ray  and radio  spectra in  the lower  panel of  Figure 4  cannot be
directly compared  with each other,  a crude estimate of  the expected
optical brightness  can 
be  obtained by connecting the  radio and
X-ray  points.  This yields  
$\sim 3\times 10^{-32}$ erg cm$^{-2}$  s$^{-1}$
Hz$^{-1}$ arcsec$^{-2}$,  i.e.,  about 
3 magnitudes below
our measured optical  upper limit. 

To summarize, we  conclude that it
is very likely  that just a slightly deeper  optical observation would
allow one to detect the  outer/inner arc (and other bright elements of
the inner PWN),  while 
much deeper optical observations  are needed to
detect the emission from the Vela PWN at large.  However, owing to the
presence  of background/foreground emission  from the  SNR as  well as
numerous bright stars in the field, longer exposures would not help to
detect the optical  PWN --- neither its X-ray-bright  central part nor
the radio-bright outskirts.  To get rid of the contaminating emission,
one should observe the field at UV wavelengths, where the radiation of
most of the field stars  would be much dimmer.  Observing in polarized
light  appears even  more promising.   Since the  synchrotron emission
from the  PWN should  be highly polarized  (as confirmed by  the radio
observations), polarimetry  observations should allow  one to minimize
the contamination  from field sources  and provide a clean  PWN image.
Polarimetry observations  of the Vela pulsar field  have been recently
obtained  by  Wagner and  Seifert  (2000) but  they  did  not yield  a
conclusive   result.   Deeper,  higher-resolution,   observations  are
needed to unveal extended polarized emission from the PWN.

\begin{figure*}
\centerline{\hbox{\psfig{figure=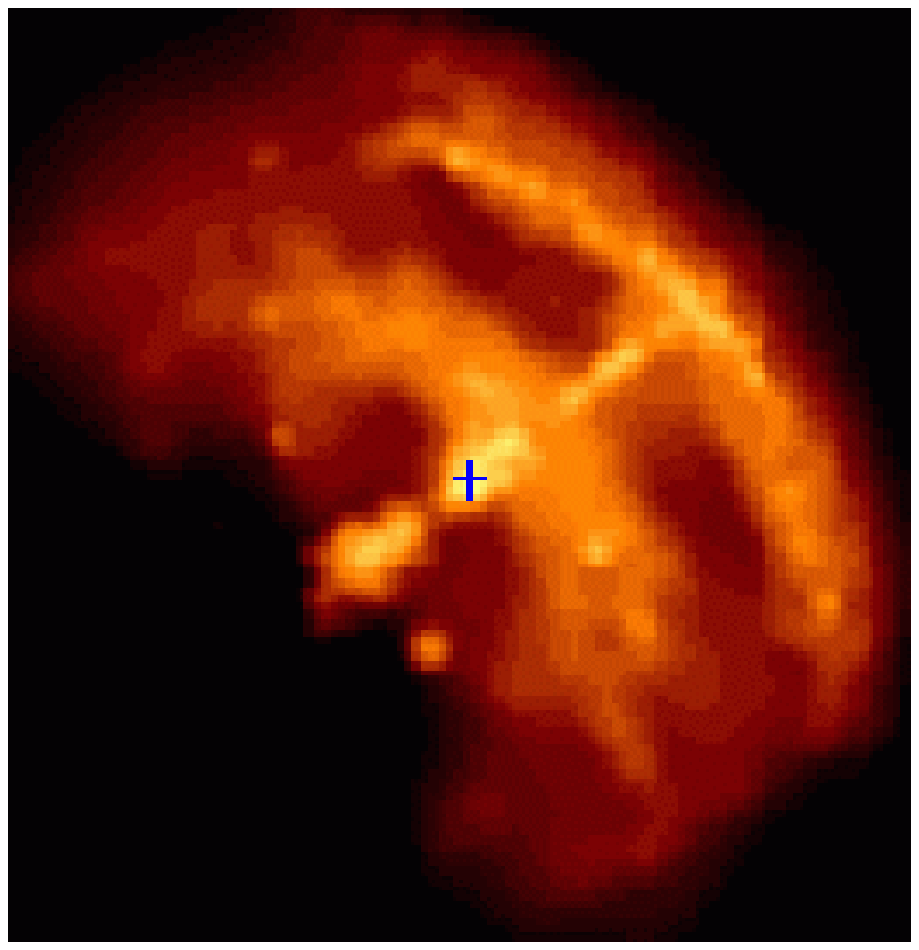,height=9.3cm,angle=0,clip=}}}
\centerline{\hbox{\psfig{figure=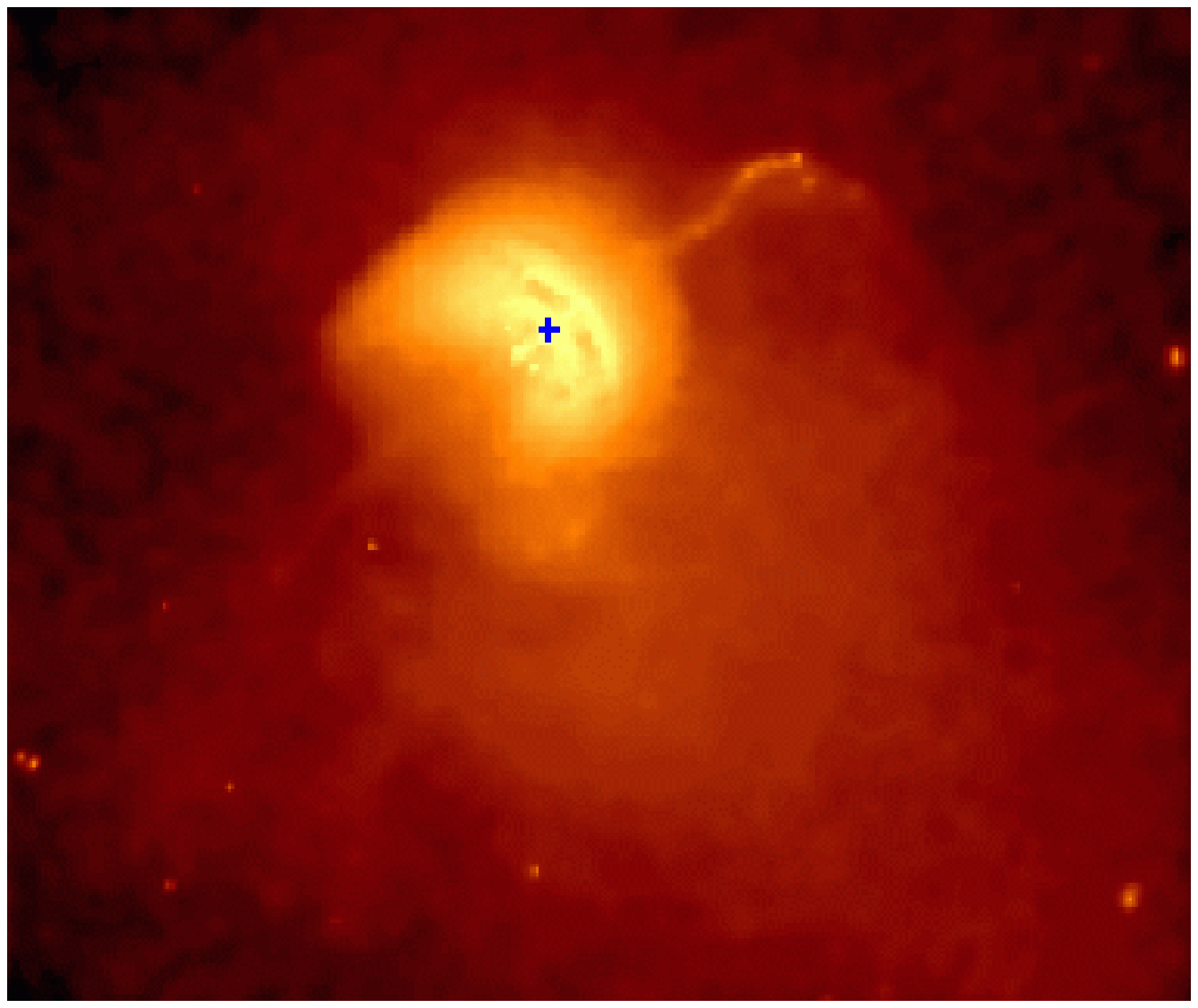,height=9.3cm,angle=0,clip=}}}
\caption{X-ray images  of the Vela  PWN, combined of  seven individual
\chan\ ACIS images (see Pavlov et al.\ 2003).   The
upper  panel  ($1\farcm60\times 1\farcm35$) shows the  bright inner PWN
around the Vela pulsar (marked by the cross), with the outer and inner
arcs, northwest  inner jet, and  southeast counter-jet. The brightness
scale in the lower panel ($6\farcm3\times5\farcm4$) was chosen so as to
show the  faint nebular structures  --- e.g., the northwest  outer jet
and the dim extended emission  southwest of the bright inner PWN.   
North  is up, East to the left.  }
\end{figure*}

\begin{figure*}
\centerline{\hbox{\psfig{figure=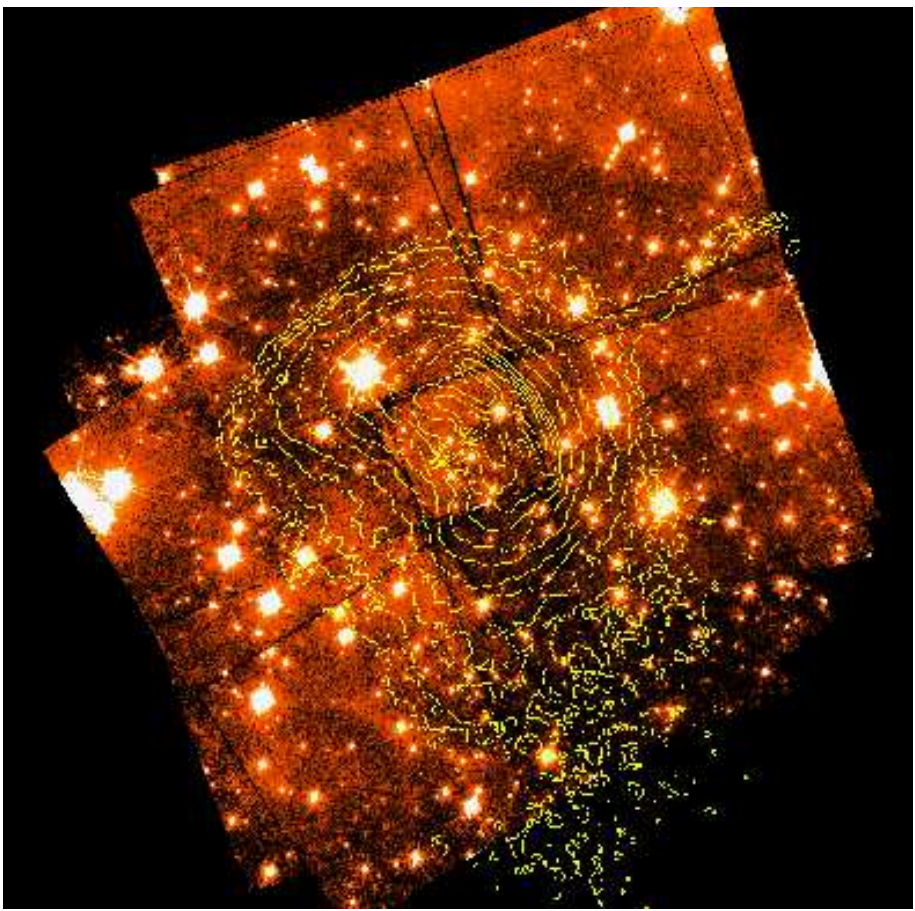,height=12cm,clip=}}}
\caption{Image of the Vela pulsar field  obtained by combining all the
WFPC2 $555W$ observations listed in Table 1 (North to  the top, East to
the  left). The  gaps  among different   CCD chips  are  evident.  The
overall integration time on the central part,  corresponding to the PC
field of  view,  is 11\,800   s   (see text).   The overlayed   contours
(logarithmic scale) correspond to   the X-ray intensity maps  obtained
from  the {\sl Chandra} ACIS  image of  the  field 
in the energy band 1--8 keV.  The  point source within  the innermost  X-ray contour is
the optical  counterpart of the Vela pulsar.  } 
\end{figure*}

\begin{figure*}
\centerline{\hbox{
\psfig{figure=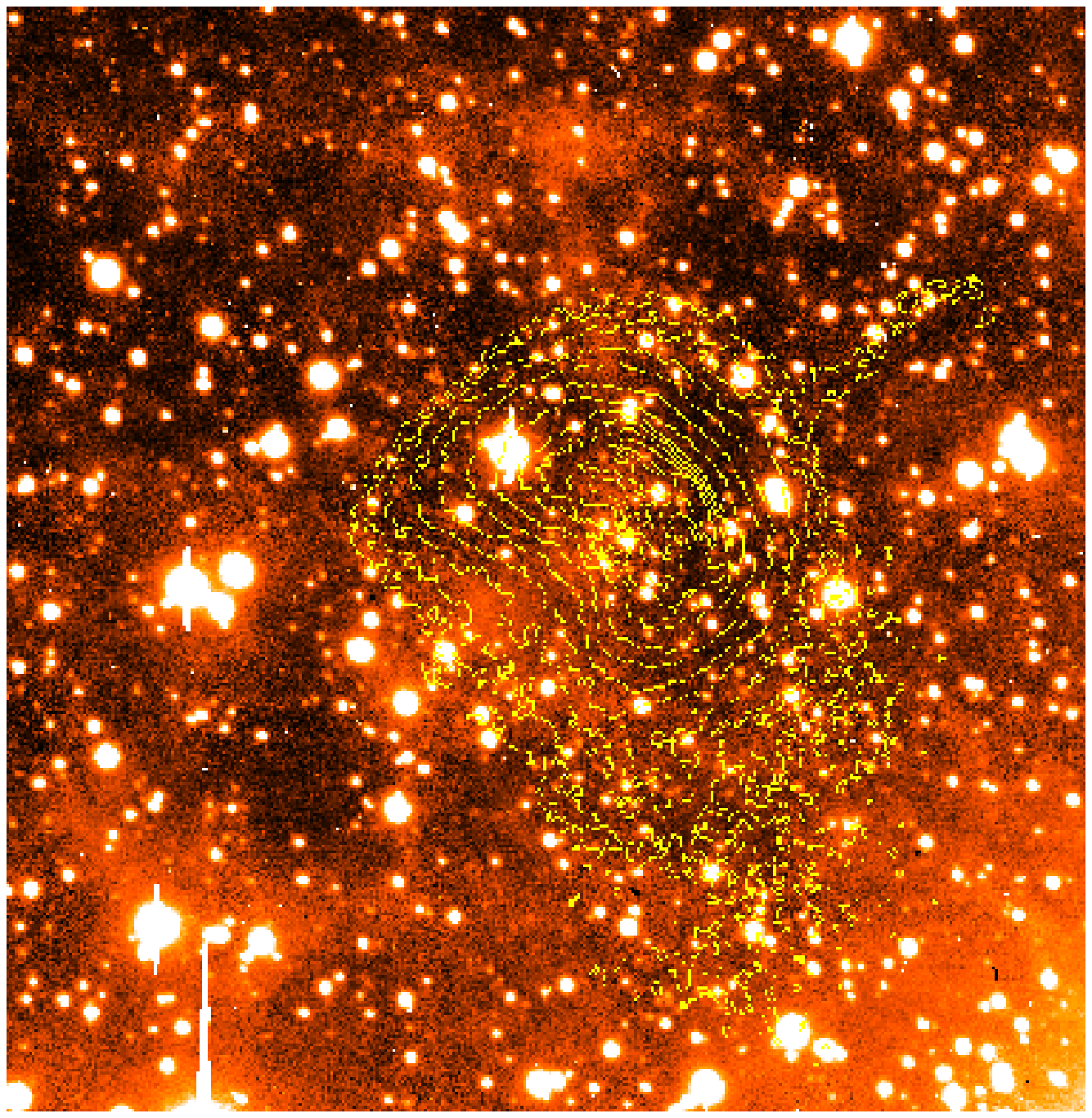,height=10cm,clip=} }}
\vspace{0.5cm}
\centerline{\hbox{
\psfig{figure=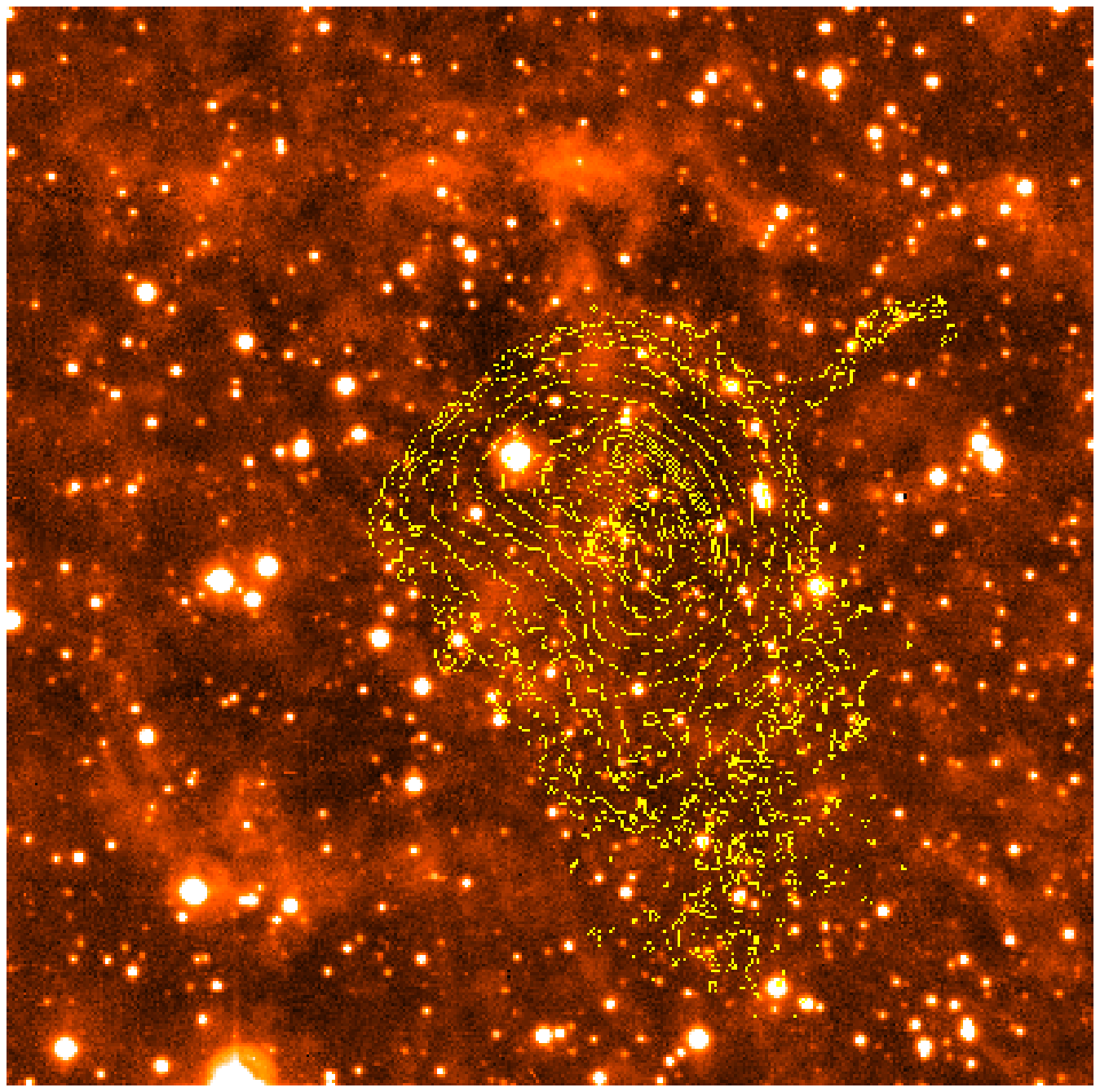,height=10cm,clip=}
}}

\caption{The upper panel shows the  combined $UBVR$ image of  the Vela
pulsar field  obtained from the {\sl NTT}/EMMI observations  listed in Table
1.  The  lower panel shows  the {\sl ESO/2.2m}  H$_{\alpha}$ image.  In both
cases the image  size is $\approx 4'\times4'$ .  North  is up, East to
the left.   The X-ray contour plots  of the Vela PWN  derived from the
seven ACIS observations (\S 2.1) are overlayed.  }

\end{figure*}

\begin{figure*}
\begin{center}
\psfig{figure=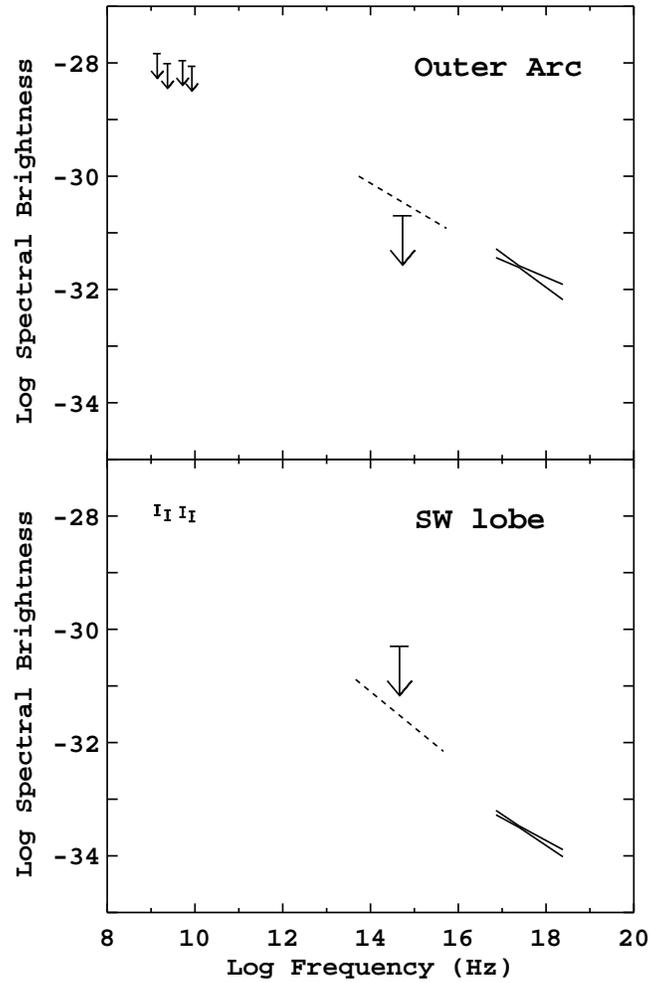,height=9cm,angle=90,clip=}
\end{center}
\caption{Spectra of
surface brightness (in erg cm$^{-2}$ s$^{-1}$ Hz$^{-1}$ arcsec$^{-2}$)
in X-rays (solid lines) and radio (points), together with the optical
upper limits, for the inner/outer arc (upper panel) and diffuse
emission southwest of the pulsar (lower panel). Expected brightness
levels in optical, based on extrapolations of the X-ray and radio data, are shown
with dashed lines. }
\end{figure*}

\begin{deluxetable}{lllcccc} 
\tablecolumns{7} 
\tablewidth{0pc} 
\tablecaption{Available optical datasets for the Vela pulsar field. }
\tablehead{ 
\colhead{Date}    & \colhead{Telescope}  &  \colhead{Instr.} & \colhead{Filter}   & \colhead{$\lambda~(\Delta \lambda)$}    & \colhead{Exp.} & \colhead{Ref.}    }
\startdata 
Jan 1995  & {\sl NTT} & EMMI-B & $U$    & 3542\AA\ (542\AA)  & 4800  & (1) \\
Jan 1995  & {\sl NTT} & EMMI-B & $B$    & 4223\AA\ (941\AA)  & 1800  & (1) \\
Jan 1995  & {\sl NTT} & EMMI-R & $V$    & 5426\AA\ (1044\AA) & 1200  & (1) \\
Jan 1995  & {\sl NTT} & EMMI-R & $R$    & 6410\AA\ (1540\AA) & 900   & (1) \\ \hline
June 1997 & {\sl HST}  & WFPC2  & $555W$ & 5500\AA\ (1200\AA)  & 2600  & (2)  \\
Jan 1998  & {\sl HST}  & WFPC2  & $555W$ & -                   & 2000  & (2) \\
June 1999 & {\sl HST}  & WFPC2  & $555W$ & -           & 2000  & (2)\\
Jan 2000  & {\sl HST}  & WFPC2  & $555W$ & -           & 2600  & (2) \\
Jul 2000  & {\sl HST}  & WFPC2  & $555W$ & -           & 2600  & (2) \\
Mar 2000  & {\sl HST}  & WFPC2  & $675W$ & 6717\AA\ (1536\AA) & 2600  & (3) \\
Mar 2000  & {\sl HST}  & WFPC2  & $814W$ & 7995\AA\ (1292\AA) & 2600  & (3)  \\ \hline
Apr 1999  & {\sl VLT} & FORS1  & $R$    & 6750\AA\ (1500\AA) & 300   & \\
Apr 1999  & {\sl VLT} & FORS1  & $I$    & 7680\AA\ (1380\AA) & 300   & \\ \hline
Apr 1999  & {\sl 2.2m} & WFI  & H$_{\alpha}$ & 6588\AA\ (74.3\AA) & 3600    &
\enddata 

\tablecomments{
First column lists the epoch of observation. 
Second and third columns show  the telescope and  the detector used  for the
observations,  respectively. The  filter  names are  listed in  column
four,  with their  pivot wavelengths  and widths  in column  five. The
total integration time per observation (in seconds) is given in column
six. The last column provides  the references: (1) Nasuti et al.\ 1997;
(2) Caraveo et al.\ (2001a); (3) Mignani \& Caraveo (2001).}
\end{deluxetable}

\begin{deluxetable}{llcccccc} 
\tablecolumns{5} 
\tablewidth{0pc} 
\tablecaption{$3\sigma$ upper limits
to the surface optical brightness of the  X-ray PWN  structures.    }
\tablehead{ 
\colhead{}  & \colhead{} &  \colhead{}    &
\multicolumn{2}{c}{Observed} &  \colhead{} &  \multicolumn{2}{c}{Extinction-corrected\tablenotemark{1}} \\ 
\cline{4-5} \cline{7-8} \\ 
\colhead{Telescope} & \colhead{Instrument} & \colhead{Filter} & \colhead{mag~arcsec$^{-2}$}   & \colhead{Flux\tablenotemark{2}} &\colhead{}& \colhead{mag~arcsec$^{-2}$} &
\colhead{Flux\tablenotemark{2}}    }
\startdata 
{\sl HST} & WFPC2  & $555W$ & 28.1 & 0.21 && 27.9 & 0.25  \\
          &        &      & 28.5--28.0  & 0.15--0.23 && 28.3--27.8 & 0.18--0.28  \\
          & WFPC2  & $675W$  & 27.5 & 0.55  && 27.3 & 0.66  \\
          &        &       & 27.9 & 0.38  && 27.7 & 0.46  \\
          & WFPC2  & $814W$  & 27.7 & 0.65  && 27.6 & 0.71  \\  
          &        &       & 28.1 & 0.45  && 28.0 & 0.49  \\  \hline
{\sl NTT} & EMMI-B & $U$   & 26.4 & 0.52  && 26.1 & 0.69  \\
          &        & $B$   & 27.4 & 0.47  && 27.1 & 0.62  \\
          & EMMI-R & $V$   & 27.1 & 0.53  && 26.9 & 0.64  \\
          &        & $R$   & 26.7 & 0.59  && 26.5 & 0.71  \\ \hline 
{\sl VLT} & FORS1  & $R$   & 27.0 & 0.45  && 26.8 & 0.54  \\
          &        & $I$   & 26.1 & 0.81  && 26.0 & 0.90  \\ 
\enddata 

\tablenotetext{1}{for $A_V=0.2$}
\tablenotetext{2}{flux values are in units of $10^{-30}$~ ergs
cm$^{-2}$ s$^{-1}$  Hz$^{-1}$ arcsec$^{-2}$} 

\tablecomments{ 
The upper limits are computed for an area of 10 arcsec$^2$. 
For the {\sl HST} results,
the  first and second rows, for a given filter, 
are the upper limits
measured in  the PC  and 
WFC chips,  respectively. Due  to the
uneven  exposure map  of  the combined  WFPC2 $555W$  image (Fig.\ 2),
slightly different upper limits are derived across the WFC field.  For
both the  {\em NTT} and {\em  VLT}, the upper limits  apply to the
overall X-ray nebula.
}

\end{deluxetable} 

\pagebreak

\begin{acknowledgements}
We warmly thank Thomas Augustejin for kindly observing for us with the
ESO/MPG 2.2m in La Silla.  We are indebted to Richard  Dodson for providing the
results of the radio observations of the Vela PWN prior to publication
and acknowledge useful  duscussions with  Divas  Sanwal.  The work  of
O.K. and   G.G.P. was partially supported  by  SAO grant G02-2071X and
NASA grant NAG5-10865. ADL thanks ASI for a fellowship. 
\end{acknowledgements}


\begin{references}

\reference {} Biretta, J.A. et al. 2002, WFPC2 Instrument Handbook, Version 7.0
 (Baltimore, STScI)

\reference {c01a} Caraveo, P. A., De Luca A., 
Mignani R. P., \& Bignami, G. F. 2001a, \apj, 561, 930

\reference{c01b} Caraveo, P. A., Mignani, R. P., De Luca, A., Wagner, S.,
 \& Bignami, G. F.  2001b, in Proc. STScI Symp., A Decade of HST Science, 
eds. M. Livio, K. Noll, \& M. Stiavelli (Baltimore: STScI), 105

\reference{c03} Caraveo, P. A., Mignani, R. P., De Luca, A., Wagner, S. \& Bignami, G. F.,  2003, in preparation 


\reference{cw00} Casertano, S. \& Wiggs, M.S.2000, see WFPC2 Instrument Handbook v5.0, 2000, ed.STScI


\reference{dodson03} Dodson, R., Lewis, D., McConnel, D., Deshpande, A.\ A. 2003,
MNRAS, submitted (astro-ph/0302373)

\reference{h85} Harnden, F. R., Grant, P. D., Seward, F. D., \& Kahn, S. M., 1985, \apj, 299, 828

\reference{hgh01} Helfand, D. J., Gotthelf, E. V., \& Halpern, J. P. 2001,
\apj, 556, 380

\reference{h95} Hester, J.  J., et al. 1995, \apj, 448, 240

\reference{} Hester, J.  J., Mori, K., Burrwos, D., et al. 2002, \apjl, 577, 49

\reference{} Hook, R. N. \& Fruchter, A. S.,
1997,  in  ASP Conf.  Series,  Vol.  125,  Astronomical Data
Analysis Software and  Systems VI, ed. G.  Hunt and H. E.  Payne
(San Francisco: ASP), 147

\reference{} Kargaltsev, O., Pavlov, G. G., Sanwal, D., \& Garmire, G. P. 2002,
in Neutron Stars in Supernova Remnants, ASP Conf.\ Ser., v.271,
eds.\ P.\ O.\ Slane \& B.\ M.\ Gaensler (ASP: San Francisco), 181

\reference{} Kennel, C. \& Coroniti, F. 1984, \apj, 283, 694

\reference{} Lewis, D., Dodson, R.,  McConnell, D., \& Deshpande, A. 2002,
in Neutron Stars in Supernova Remnants, ASP Conf.\ Ser., v.271,
eds.  P. O. Slane \& B. M. Gaensler (ASP: San Francisco), 191


\reference{} Markwardt, C. B. \&
 \"Ogelman, H. B. 1998, Mem.\ Soc.\ Astr.\ It., vol.69, p.927


\reference{} Mignani, R. P. \& Caraveo, P. A., 2001, \aap , 376, 213

\reference{} Monet, D. G., et al. 1998, USNO-A20 (Washington: US Nav. Obs.)

\reference{} Nasuti, F. P., Mignani, R., Caraveo, P. A. \& Bignami, G. F. 1997,  \aap,  323, 839

\reference{} \"Ogelman, H. B, Koch-Miramond, L., \& Aurie\'ere, M. 1989, \apjl,  342, 83

\reference{} \"Ogelman, H., Finley, J. P., \& Zimmerman, H. U. 1993, \nat, 361, 136

\reference{} Pavlov, G. G., Sanwal, D., Garmire, G. P., Zavlin, V. E., Burwitz, V., \& Dodson, R. G. 2000, AAS Meeting 196, \#37.04

\reference{} Pavlov, G. G., Zavlin, V. E., Sanwal, D., Burwitz, V., 
\& Garmire, G. P. 2001a, \apj, 552, 129

\reference{} Pavlov, G. G., Kargaltsev, O. Y., Sanwal, D., \& Garmire, G. P.,2001b  \apjl, 554, 189

\reference{} Pavlov, G. G., Teter, M. A., Kargaltsev, O., \&
Sanwal, D. 
2003, ApJ, accepted

\reference{} Russell, J. L., Lasker, B. M., McLean, B. J., Sturch, C. R. \& Jenkner, H., 1990, \aj 99, 2059


\reference{} Scargle, J. D. 1969, \apj, 156, 401

\reference{} Wagner, S. and Seifert S., 2000, in Proc. of the
            177th IAU Colloquium Pulsar Astronomy - 2000 and Beyond,
            ASP Conference Series, Vol. 202, p. 315, M. Kramer,
            N. Wex, and N. Wielebinski Eds.

\end{references}
\end{document}